\documentclass{article}
\usepackage{fullpage}
\usepackage{amsmath}
\usepackage{amssymb}
\usepackage{orcidlink}
\usepackage{caption}
\usepackage{float}
\usepackage{newfloat}
\usepackage{graphicx}
\usepackage{rotating}
\usepackage{chngcntr}
\usepackage{placeins}
\usepackage{hyperref}

\DeclareMathOperator{\sgn}{sgn}
\DeclareMathOperator{\Var}{Var}
\DeclareMathOperator{\E}{\mathbf{E}}
\DeclareMathOperator{\Cov}{Cov}

\counterwithin*{equation}{part}
\counterwithin*{figure}{part}
\counterwithin*{table}{part}
\counterwithin*{section}{part}
\counterwithin*{subsection}{part}

\begin{document}

\renewcommand{\theequation}{\thepart.\arabic{equation}}
\renewcommand{\thefigure}{\thepart.\arabic{figure}}
\renewcommand{\thetable}{\thepart.\arabic{table}}
\renewcommand{\thesection}{\thepart.\arabic{section}}

\title{Loss Functions for Detecting Outliers in Panel Data\footnote{This paper reports the general results of research undertaken by Census Bureau Staff. The views expressed are attributable to the authors and do not necessarily reflect those of the Census Bureau.  The authors would like to thank participants in a Statistics Colloquium, Department of Applied and Engineering Statistics, George Mason University, for valuable comments.  A previous version of Part \ref{part1} was presented to that Colloquium and to the Spring 2000 meetings of the Federal-State Cooperative Program for Population Estimates, March, 2000, Los Angeles, CA. A previous version of Part \ref{part2} was also presented to the above-mentioned Statistics Colloquium. Part \ref{part2}'s graphics were created using GNU Octave.}}
\author{Charles D. Coleman\orcidlink{0000-0001-6940-8117}\thanks{ 
Corresponding author,
Timely Analytics, LLC, 
E-mail: info@timely-analytics.com}
and Thomas Bryan\thanks{Bryan Geodemographics}
}

\maketitle

\abstract{The detection of outliers is of critical importance in the assurance of data quality.   Outliers may exist in observed data or in data derived from these observed data, such as estimates and forecasts.  An outlier may indicate a problem with its data generation process or may simply be a true, but unusual, statement about the world.  Without making any distributional assumptions, we proposes the use of loss functions to detect these outliers in panel data.  

Part \ref{part1} covers nonnegative data.  We axiomatically derive an unsigned loss function.  We then develop a signed loss function ito account for positive and negative outliers separately.  In the case of nominal time we obtain an exact parametrization of the loss function.  A time-invariant loss function permits the comparison of data at multiple times on the same basis.  We provide several examples, including an example in which the outliers are classified by another variable.

Part \ref{part2} covers data of mixed sign.  Similar to Part \ref{part1}, we axiomatically develop unsigned and signed loss functions. We search for optimal values of the loss function parameter using graphs.}

\section*{Introduction}

In general, an outlier is an observation which departs from the norm (however defined) in a set of observations.  Outliers can indicate a problem with the data generation process (i.e., anomalies) or may be true, but unusual, statements about reality.\footnote{This is similar to Hoaglin’s (1983, p. 39-40) use of “outside cutoffs” to identify “outside values.”}   Our purpose in detecting outliers is to determine whether there are problems with the data generation process(es).  In terms of Barnett and Lewis (1994, p. 37), we are testing for discordancy. This paper specializes the problem of detecting outliers to panel data, such as estimates and forecasts.  Panel data are cross-sectional time series, such as a time series of population or migration estimates for a set of areas.\footnote{The bidimensionality of data searched for outliers is not unique:  DuMouchel (1999), Albert (1997), and Rudas, Clogg and Lindsay (1994) search for outliers in contingency tables.  The contingency table approach differs in that time need not be a dimension and that parametric assumptions are made.}  We make no distributional assumptions, as we assume the data generation processes to be unknown and nonidentical.\footnote{This obviates the use of parametric techniques, in which observations are tested for departure from a predetermined, hypothesized distribution.}  Time may be either chronological or nominal.  Nominal time indexes different sets of estimates or forecasts for the same cross-sectional units and chronological time.  Time is nominal in this context because the different estimates sets have no natural ordering.  Comparing cross-sectional estimates to their true values is an instance of nominal time. Nominal time is only applicable to the case of nonnegative data due to the importance of geometry in the mixed-signs case, so we confine its discussion to Part \ref{part1}.  Part \ref{part1} covers nonnegative data.  Part \ref{part2} relaxes the sign restriction to examine data of mixed sign.  In each of these Parts, we specify the assumptions and create the simplest loss function that satisfies those assumptions and admits the property that the loss for a given relative difference (properly defined) increases in the initial value.  A Lie symmetry permits elimination of an exponent.  This enables each value of the remaining exponent to be associated with pairs of exponents  lying on a line for the equivalent decompositions into power of the absolute difference and absolute relative difference (properly defined).

Part \ref{part3} concludes this article.

\part{Nonnegative Data\label{part1}}

\section{Introduction to Part \ref{part1}}

Stock data, such as population, bank deposits and inventories are normally nonnegative data. Some flow data, such as births and deaths are also nonnegative. 
We develop loss function in Section \ref{LF1} to compare two sets of nonnegative values.  We create four types of loss functions: unsigned, signed, time-invariant and a special case when the data are generated by processes sharing a particular set of criteria.  Section \ref{applications} discusses some applications, including general usage of loss functions, parametrizing loss functions from preexisting outlier criteria, using loss functions with GIS and parameterizing a variant of the loss functions of Section  \ref{LF1} when the outliers are classified by another variable.  These examples are based on actual Census Bureau applications.

\section{The Loss Function\label{LF1}}

We describe the assumptions used to generate the loss function $L(F;B)$ and its variants, where $F$ is the future value and $B$ is the base period value.  The loss function is the penalty associated with the difference between $F$ and $B$.  Roughly speaking, the greater the difference between $F$ and $B$, the greater the loss.  Initially, $F$ is assumed to be one period after $B$.  After we make the necessary assumptions in Subsection \ref{uLF1}, we specify the simplest forms of $L$ that also allow the values of the parameters of $L$ that make it increase in $B$ for a given relative difference. We then give the condition for the latter to happen.   In Subsection \ref{timeinvariant} we generalize $L$ to situations in which $F$ and $B$ may not be exactly one period apart.  In Subsection \ref{sLF1} we introduce the signed loss function for cases in which the sign of the difference is an additional important criterion. In Subsection  \ref{twosets1} we parametrize $L$ for comparing two sets of estimates of the same parameters.  Throughout this paper, we assume $B$ and $F$ to be positive.  Zeros, which frequently arise in the contexts in which we have worked, are either recoded to small values or omitted from the analysis.

\subsection{The Unsigned Loss Function\label{uLF1}}

We construct the unsigned loss function $L$ by specifying three assumptions.\footnote{The exposition in this Subsection through Property 1 (except for the use of Lie symmetry) follows Coleman (2025).} The first assumption is that $L$ is symmetric in the differences:\newline
{\bf Assumption I.1:} $L(B + \epsilon,B) = L(B - \epsilon,B)$  for all $B,\epsilon \ge 0$.\newline
This assumption is not as innocuous as it looks.  It is quite possible that, at least for some range of $B$, that positive and negative differences have differential impacts. However, the resulting asymmetry complicates the definition of $L$.   We relax this assumption in Subsection \ref{sLF1} by developing the signed loss function, which allows the possibility of incorporating the direction of the difference $\epsilon$.  The symmetry of $L$ allows us to use the equivalent notation $\ell (\epsilon,B) = L(F;B)$ where $\epsilon = |F - B|$.

The next assumption makes $L$, or, equivalently, $\ell$, increasing in the difference $\epsilon$:\newline
{\bf Assumption I.2:} $\partial \ell / \partial \epsilon > 0$ for all $\epsilon \ge 0$.\newline
Note that this assumption is stated in terms of $\ell$, rather than $L$.  This assumption is quite intuitive, as it states that smaller differences are preferred to larger ones.

Finally, we want $L$, or, equivalently,   $\ell$, to decrease in $B$.  This means that for a given value of $\epsilon$, the loss associated with it decreases with its associated initial value.  This has two justifications.  First, for example, a difference of 500 when the initial value is 1,000 is a whopping 50\%, a highly significant difference.  However, the same difference, when the initial value is 1,000,000 is akin to a roundoff error.  Second, when performing estimates or taking samples, the coefficient of variation  $\sigma / \mu$, where $\sigma$ is the standard deviation and $\mu$ is the expected value, decreases in $B$. Coleman's experience is that all areas tend to have about the same roundoff errors.  Again, these are proportionately greater in small areas.  We state this formally as:\newline
{\bf Assumption I.3:} $\partial \ell / \partial B < 0$, or, equivalently, $\partial L/ \partial B < 0$  for all $B > 0$.

The simplest function which satisfies Assumptions I.1–I.3 and admits Property I.1 below is the Cobb-Douglas function
\begin{subequations}
	\begin{equation}
		\epsilon^p B^q
           \end{equation}
or, equivalently,
	\begin{equation}
		|F-B|^p B^q.  \label{FBloss}
           \end{equation}
\end{subequations}

An observed pair $(F;B)$ is an outlier whenever $L(F;B) > C$, where $C$ is a predetermined critical value.\footnote{$C$ can also be determined from the data by taking a predetermined quantile or a multiple of the interquartile range of $L$ (Tukey 1977).}   We  refer to outliers as being critical.  We  also refer to the equation $L(F;B) = C$ as the equation of criticality.  The choice of $p, q$ and $C$ is an empirical matter.   Only a practitioner’s experience with data can determine when data are suspect and incorporate these suspicions into parameters.   One thing to note is that the loss function is ordinal: raising $L$ or $\ell$ and $C$ to any positive power $m$ leaves the rankings of losses unchanged.  It is only the rankings of losses that are important.   Formally, the transformation $p \to p_m = mp$, $q \to q_m = mq$ and $C \to C_m = mC$ creates a new equation of the same form.  The same independent variables create the same criticality (or lack thereof).  This is called a Lie symmetry or Lie invariance.  The upshot is that one parameter can be fixed and the other two varied to generate new loss functions and critical values.   In subsequent development, we set $p$ equal to 1.  Thus,
\begin{subequations}
	\begin{equation}
		\ell(F,B) = \epsilon B^q \label{ell1}
           \end{equation}
or, equivalently,
	\begin{equation}
		L(F;B) = |F-B| B^q.  \label{L1}
           \end{equation}
\end{subequations}

A desirable property of the loss function is that it can be made to rise in $B$ for a given absolute relative difference.  The absolute relative difference is: 
\begin{equation}
	|F-B| B^{-1}
\end{equation}
Note that, in this case, $q = -1$.  Choosing $q > -1$ makes the loss function rise in $B$, for a given absolute relative difference.  We state this as\newline
{\bf Property I.1:}
The loss function defined by equations (\ref{ell1}) and (\ref{L1}) increases in $B$ for any given absolute relative difference.  This is assured whenever $q > -1$.

The reader may notice that $q = 0$ turns the first terms of equations (\ref{ell1}) and (\ref{L1}) into the absolute values of the differences.  Thus, values of $q$ between 0 and $–1$ represent various tradeoffs of absolute differences and absolute relative differences.  Consider the product of the $r$th power of the absolute difference and the $s$th power of the absolute relative difference, where $r, s > 0$, 
\begin{equation}
|F - B|^r \left(\frac{|F-B| }{B}\right)^s.\label{lossprod}
\end{equation}
By the Lie symmetry invoked above, loss function (\ref{lossprod}) is isomorphic to the loss function  $|F-B|B^{-s/r+s}$.  Thus, any value of $q$ corresponds to an infinite number of pairs $(r, s)$ where $q = –s / (r + s)$.  Geometrically, the same loss function is generated for all $(r, s)$ lying on the line $r  =  –(1 + q) s$.

Subsection \ref{DR} provides an example of a variant of this loss function approach when the outliers are classified by another variable.  Changes in the variable of interest are declared outliers contingent on another variable.  Assumption I.3 is violated, but, in this example, Property I.1 holds.

\subsection{The Time-Invariant Loss Function\label{timeinvariant}}

Instead of considering the single set of future data,  $\mathbf{F} = \{F_i\}_{i=1}^n$, where $i$ indexes the $n$ observations, consider the sets  $\mathbf{F}_t = \{F_{it}\}_{i=1}^n$, where $t$ is the amount of time elapsed since the base date.  We wish to develop a loss function which allows us to make comparisons across time on the same basis, by explicitly incorporating $t$ into the loss function.  One way of incorporating time-invariance is to use the geometric average absolute relative change.  Since the base date is constant for all observations, we may rewrite $L$ as
\begin{equation}
L(F_{it};B_i,t)=\left(\frac{|F_{it}-B_i|}{B_i}\right)^\frac{1}{t} B_i^{q+1} = |F_{it}-B_i|^\frac{1}{t} B_i^{q+1+\frac{1}{t}}\label{Lit}
\end{equation}
Equation (\ref{Lit}) puts the geometric average absolute relative difference on the same basis for all $t$.

Now, we can use Lie symmetry to make the exponent of  $|F_{it}-B_i|$ unity by multiplying both exponents by $t$. 
\begin{equation}
L(F_{it};B_i,t)=  |F_{it}-B_i| B_i^{tq+t-1}.
\end{equation}
The reader can verify that $–1 < tq + t – 1 < 0$ for $0 < t \le 1$ when $0 > q > –1$.  Thus, the data times should be rescaled so that the last one becomes 1.

 \subsection{The Signed Loss Function\label{sLF1}}

At times, not only the value of the loss function is important, but also the sign of the difference.  Different outlier generation processes may manifest themselves by producing predominantly positive or negative differences.  We can account for these by creating the signed loss function $S$, which is simply the loss function $L$, multiplied by the signum function of the difference:
 \begin{equation}
	S(F;B) = |F-B| B^q \sgn)(F-B) = (F-B) B^q, \label{S}
 \end{equation}
where $\sgn x = +1$ for $x > 0$, 0 for $x = 0$, and $–1$ for $x < 0$.

Using $S$, we can create different critical values for loss, depending on whether the difference is positive or negative.  To wit, one can pick $C_+, C_-, C_+ \ne -C_-$, such that a pair $(F, B)$ is declared an outlier if either $S(F;B) < C_-$ or $S(F;B) > C_+$.  Again, the choice of whether to use $S$ and then use asymmetric critical bounds is an empirical matter.   For example, since, by assumption, negative values of $F$ are impossible, then asymmetric critical bounds and/or parameters may be necessary to detect cases in which $F$ becomes very small relative to $B$.

The time-invariant signed loss function  is
 \begin{equation}
S(F_{it};B_i,t)=  (F_{it}-B_i) B_i^{tq+t-1}.
\end{equation}

\subsection{Comparing Two Sets of Data: A Specialization of the Loss Function\label{twosets1}}

Suppose that the sets $\mathbf{B} = \{B_i\}$ and $\mathbf{F} = \{F_i\}$  represent two versions of estimates of the true values  $\mathbf{A} = \{A_i\}$.\footnote{The derivations in this Subsection were done with assistance from ChatGPT 5.}  Suppose that both the $B_i$ and $F_i$ are unbiased estimators of the $A_i$ and that their variances are proportionate to the $A_i$  (i.e., $\Var(B_i) = \Var(F_i) = \sigma^2A_i$.)\footnote{ This can be justified by assuming that the $B_i$ and $F_i$ are both sums of $A_i$ independent (or, more weakly, jointly uncorrelated) estimators with expected value 1 and variance $\sigma^2$.}  Suppressing subscripts, consider the expected value of the squared difference of $B$ and $F$, $\E(F – B)^2$:
\begin{equation}
\E(F-B)^2=\Var(F)+\Var(B)-2\Cov(F,B)+(\E F-\E B)^2
\end{equation}
Since $\E B = \E F = A$  the last term is zero and is thus dropped:
\begin{equation}
\E(F-B)^2=\Var(F)+\Var(B)-2\Cov(F,B).
\end{equation}
Substituting for the variances obtains:
\begin{equation}
\E(F-B)^2=2\sigma^2 A-2\Cov(F,B). \label{A2Cov}
\end{equation}
Letting $\rho$ be the correlation between $B$ and $F$ we can express $\Cov(F,B)$ as
\begin{equation}
\Cov(F,B) = \rho\Var(F)\Var(B)=\rho\sigma^2A. \label{Cov}
\end{equation}
Substitutiong equation (\ref{Cov}) into equation (\ref{A2Cov}) yields
\begin{equation}
\E(F-B)^2=2(1-\rho)\sigma^2 A \propto A.
\end{equation}
Removing the expection means than, on average,
\begin{equation}
(F-B)^2 \propto A = \E B.
\end{equation}
Again, removing the expectation and taking the square root means that, on average,
\begin{equation}
|F-B| \propto  B^\frac{1}{2}.
\end{equation}
Rearranging and dividing both sides by $B^{1/2}$ obtains 
\begin{equation}
|F-B| B^{-\frac{1}{2}} \propto 1. 
\end{equation}
This means that, in this situation, we can use the loss functions (\ref{ell1}) and (\ref{L1}) with $q = –1/2$.  Since the null distributions of $B$ and $F$ are assumed unknown, it is impossible to do any significance testing.  Moreover, since we are dealing with the entire population, sampling theory breaks down.  Even though estimates of $\rho$ and $\sigma$ can be computed from the data, it is unclear how to use normal (or any other distributional) theory for testing.  The simplest thing to do in this situation is to rank the losses in descending order and empirically decide which critical value to use.

Again, the signed loss function (\ref{S}) can be used with $q = –1/2$.

\section{Applications\label{applications}}

We illustrate the use of loss functions by first outlining a general procedure for using loss functions in Subsection \ref{general}.  Next, three different examples of loss functions are shown.  In the first example, in Subsection \ref{preexisting}, preexisting outlier criteria in terms of critical ratios by size class are transformed into a loss function. The second example, in Subsection \ref{GIS}, uses real-world data and GIS to compare the 2010 and 2020 Census county counts using the $q = –1/2$ loss function of Subsection \ref{twosets1} as well as absolute and absolute relative differences to evaluate differences between these two sets of estimates.  The third example, in Subsection \ref{DR}, uses preexisting outlier criteria for a variable, which are classified by another variable, to generate a different type of loss function.

\subsection{General Procedure for Using Loss Functions\label{general}}

Loss function evaluations usually begin by recoding zeros to a small positive value  (the exact value determined by the range and smallest value of the data) and setting $q = -.5$.  If time is chronological, the analyst then has to examine the data and the rankings of their associated losses.   If, in their opinion, too many observations with small changes occurring to small base values are ranked highly, then $q$ should be increased.  If, on the other hand, too many observations with small changes to large base values are ranked highly, then $q$ should be decreased.  This process continues until the analyst is satisfied with the loss rankings.  Our experience has found that initially changing $q$ by .1 is satisfactory.  If time is nominal, our experience is that two comparisons should be done using both variables for $B$.

Bryan (1999) suggests using $q = \log(range)/25 – 1$ as a starting point, where $range$ is the range of the data under examination.   The motivation is that as the range increases, the absolute difference should take increasing weight.  Conversely, as the range decreases, the absolute relative difference should take increasing weight.  This criterion is not scale-invariant, so it should only be used with discrete data.

\subsection{Preexisting Outlier Criteria\label{preexisting}}

Table \ref{table1} provides an example of base values $B$, grouped by size class, the ratios $\epsilon/B$, which are considered to be outliers, and the ranges of $\epsilon = |F - B|$.   The midpoints of the $B$, and $\epsilon$ ranges are also shown, for later use. Table \ref{table1} shows some peculiarities.  The ranges corresponding to $\epsilon/B = 1.50$ and 3.00 have the same minimum $\epsilon$, effectively violating Assumption I.3.   Worse, the range corresponding to $\epsilon/B = .40$ has all of its $\epsilon$ range higher than the corresponding values in the adjacent ranges.  This is in spite of a reasonable progression of values of $\epsilon/B$, which accords with Property 1.  One solution to these problems is to have the decision-maker revise the $B$ size classes and/or the values of $\epsilon/B$.  However, this may not always be possible. In these cases, the estimation technique for $\ell$ described below must be used with the ranges whose midpoints violate monotonicty discarded.

\begin{table}[H]
\centering
	\caption{$B$ and $\epsilon$, by Size Class}
\vspace*{\bigskipamount}
    \begin{tabular}{|l|r|r|r|r|}
    \hline
        $B$ Size Class & $\epsilon/B$ & $\epsilon$, range & Midpoint of $B$ & Midpoint of $\epsilon$ \\ \hline
        50,000+ & 0.05 & 2,500+ & - & - \\ \hline
        25,000-49,999 & 0.15 & 3,750-7,499 & 37,500 & 5,625 \\ \hline
        10,000-24,999 & 0.40 & 4,000-9,999 & 17,500 & 7,000 \\ \hline
        5,000-9,999 & 0.60 & 3,000-5,999 & 7,500 & 4,500 \\ \hline
        2,500-4,999 & 1.00 & 2,500-4,999 & 3,750 & 3,750 \\ \hline
        1,000-2,499 & 1.50 & 1,500-3,749 & 1,250 & 2,625 \\ \hline
        500-999 & 3.00 & 1,500-2,999 & 750 & 2,250 \\ \hline
        0-499 & 4.00 & 0-1,999 & 250 & 1,000 \\ \hline
    \end{tabular}\label{table1}
\end{table}

The two rightmost columns of Table  \ref{table1} show the midpoints of the ranges of $B$ and $\epsilon$.  These are used to construct $\ell$ by linear regression.   Essentially, these points are used to define a level curve of $\ell$  in the $(B,\epsilon)$ space.  By the Implicit Function Theorem, these points satisfy the equation
\begin{equation}
\frac{d\epsilon}{dB}=-\frac{\partial\ell / \partial B}{\partial\ell / \partial \epsilon} = -\frac{q\epsilon B^{q-1}}{B^q} = - \frac{q\epsilon}{B}. \label{IFT}
\end{equation}
Upon rearrangement, equation (\ref{IFT}) becomes
\begin{equation}
\frac{d\epsilon}{\epsilon}=-q\frac{dB}{B}.
\end{equation}
This can be integrated to obtain
\begin{equation}
\log \epsilon = -q\log B + K.\label{loglog}
\end{equation}
where $K$ is a constant of integration. Equation (\ref{loglog}) can then be estimated by linear regression to find the values of the parameters –q and K. (\ref{loglog}) is then exponentiated and rearranged to obtain
\begin{equation}
\ell(\epsilon,B) = \epsilon B^q = e^K = C.
\end{equation}
Applying this methodology to Table 1,  we obtain the estimates $-q = 0.32675$ and $K = 5.405415$.\footnote{Remember that we are omitting the range corresponding to $\epsilon/B = .40$.}  Thus, $L(F;B) = |F-B|B^{-0.32765} = \ell(\epsilon,B) = \epsilon B^{-0.32765}$ and $C = \exp(5.405415) = 222.61$.

\subsection{GIS\label{GIS}}

Population change between censuses is of great interest.  The main problem in interpreting change is addressed by Property I.1:  The coefficient of variation decreases in population.  We examine population change by county 2010--2020.  The 2010 population is the U.S. Census Bureaus' Population Estimates Program 2010 base.  The 2020 population is the population as enumerated by the 2020 census.  The Maps omit Alaska and Hawaii for convenience.  

Map I.1 displays absolute changes in population.   This map has some noteworthy features.  The first set consists of groups of low growth The first of these is the band of counties in the Western Plains with low values.  The second is a barely discernable band of counties inland of the Fall Line from Pennsylvania to Virginia.  Finally, another band of counties with low values runs along the northern borders of Tennessee and Arkansas.  All of these are interrupted by high value counties.  Their values generally reflect their low populations.  The Washington-Boston megalopolis is easily seen with its high absolute differences.  This megalopolis really extends from Richmond, Virginia to Portland, Maine.

Map I.2 displays the absolute percent differences.  The first thing we notice is less clustering of counties relative to Map I.1.  The Western Plains counties identified earlier show a great range of absolute growth rates, reflecting their small sizes.  The band of low growth counties inland of the Fall Line has grown denser and longer, reaching into Tennessee and parts of New England.  The high absolute difference areas of Map I.1 also generally have high absolute percentage growth rates: More urban counties are experience higher absolute growth rates. The Boston-Washington metropolis splits into 4 areas: an area of high absolute percentage growth from Richmond to Washington, an area of lesser absolute percentage growth around Boston and Portland, an area of even less absolute  percentage growth from Baltimore to New York City and an area of very low absolute percentage growth in central Connecticut.

Map I.3 displays the $q=-.5$ loss functions by county.  The low growth area in the Western Plains expands to Wyoming, Minnesota, Michigan, Iowa and Missouri.  While many of the counties have darker coloration than in Map I.1, that could be an artifact of the ranges.  The lines of low growth counties in the East remain, but are thinner.  The Boston-Washington megalopolis reappears with the exception of central Connecticut.  By varying $q$ and the ranges, the analyst can gain insights into county population changes.  Using a signed loss function would enable identifying growing and declining counties separately and provide better insights.  Doing these enables loss functions to become data exploration tools.

\begin{sidewaysfigure}\label{map1}
	 \includegraphics[width=9in]{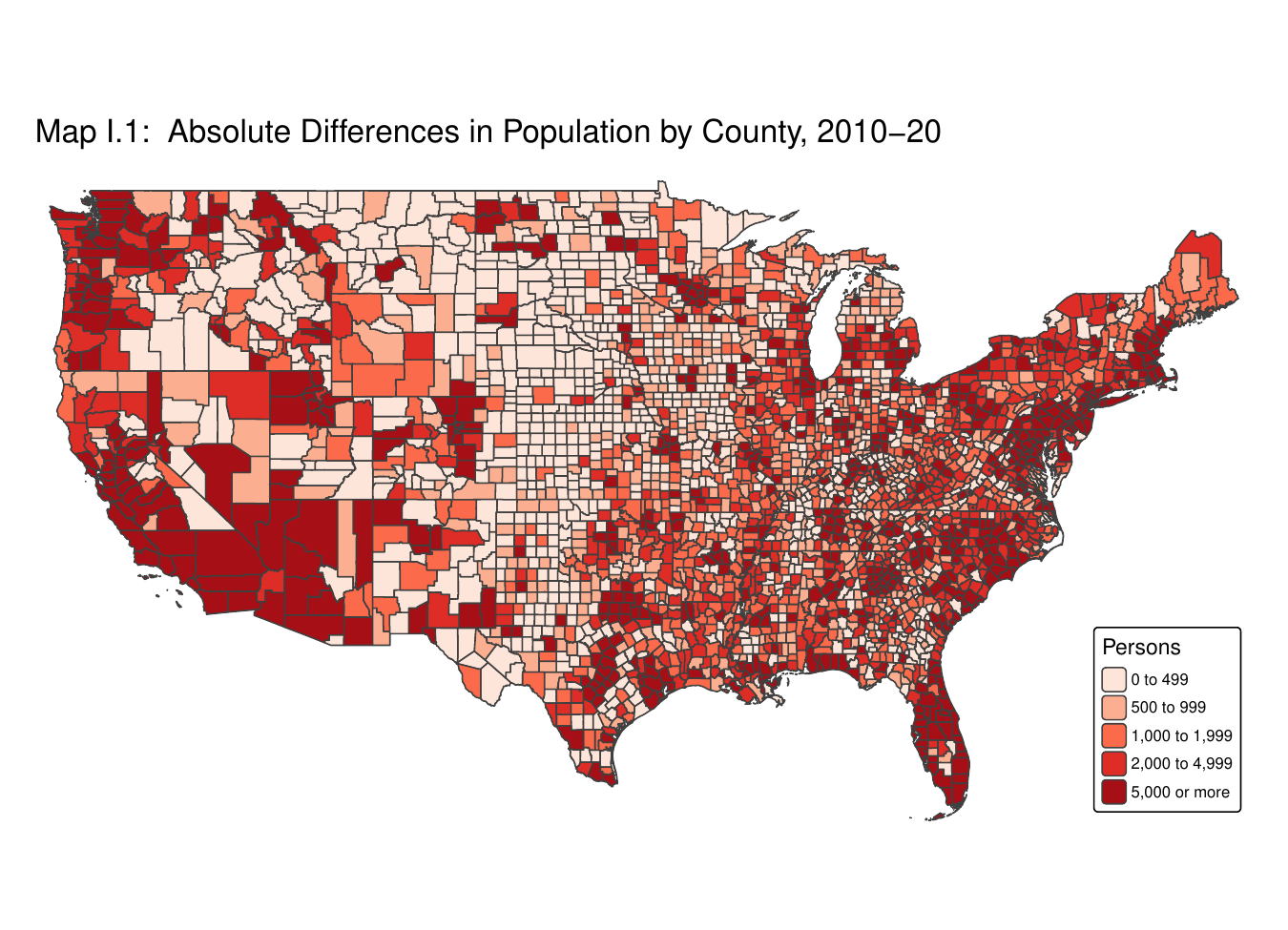}
\end{sidewaysfigure}

\begin{sidewaysfigure}\label{map2}
	 \includegraphics[width=9in]{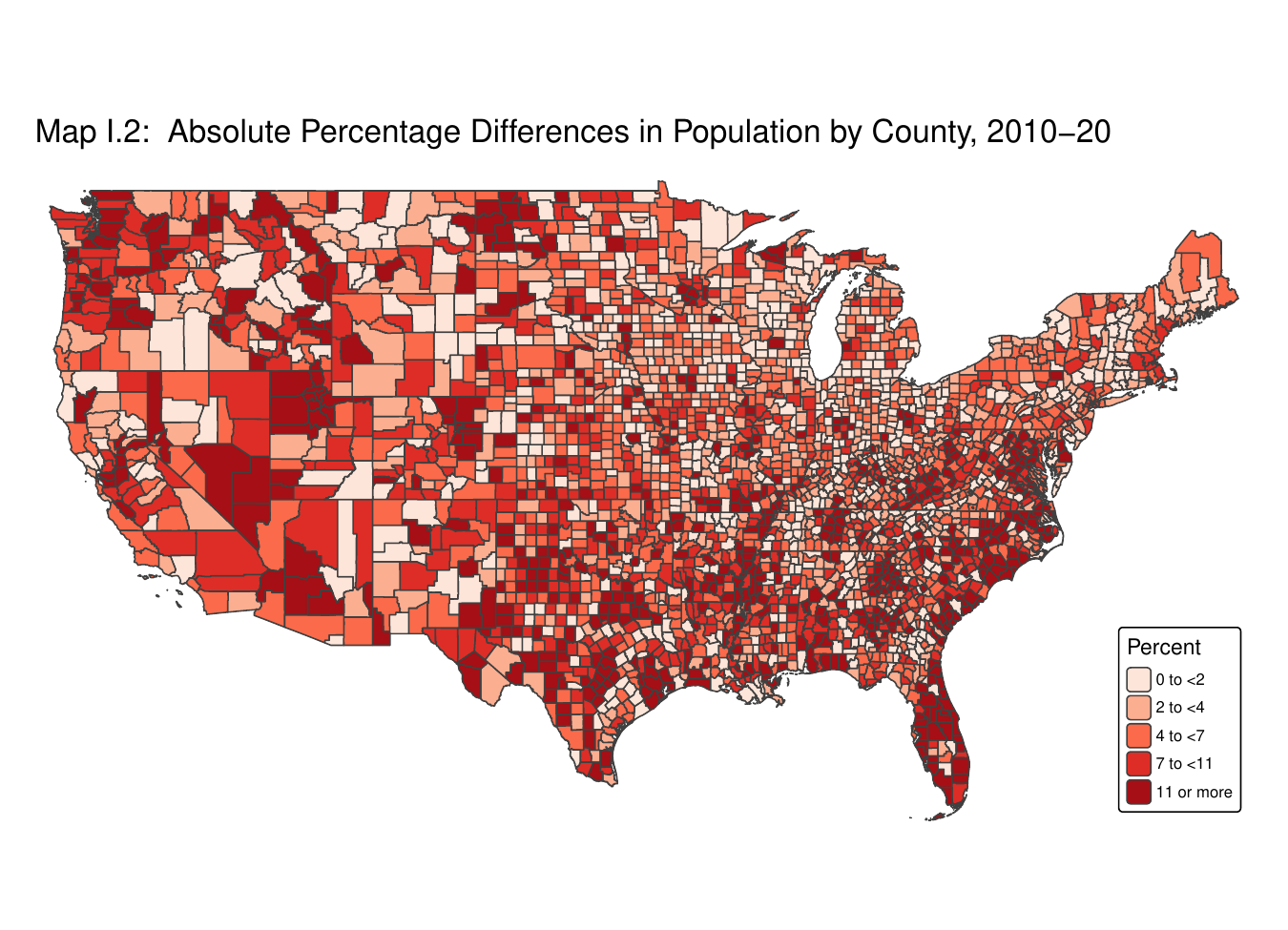}
\end{sidewaysfigure}

\begin{sidewaysfigure}\label{map3}
	 \includegraphics[width=9in]{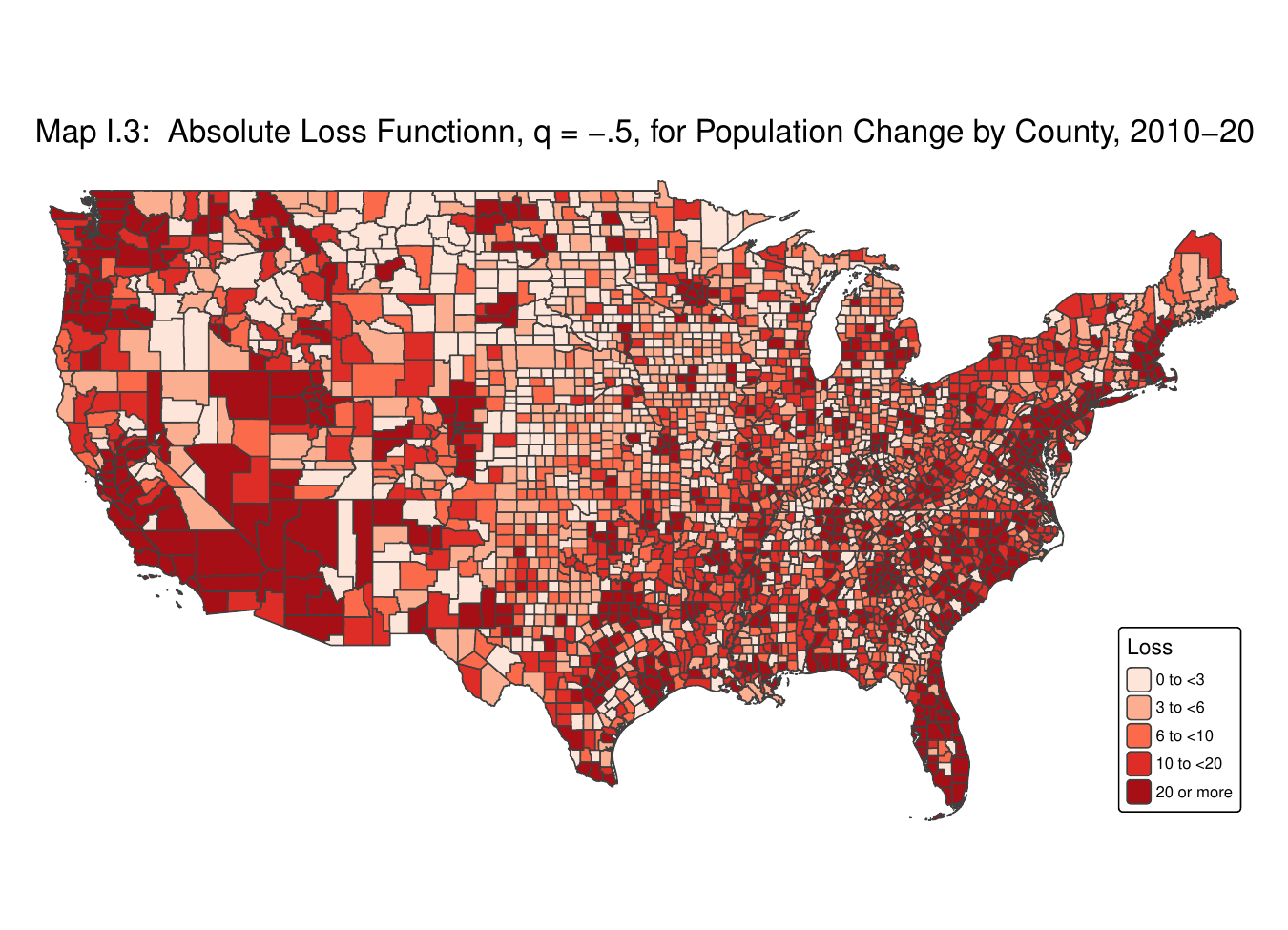}
\end{sidewaysfigure}

\subsection{Example with Outliers Classified by Another Variable\label{DR}}

\begin{figure*}
  \caption{$D$ as a Function of $R$}\label{DRfig}
 \centering
 \includegraphics{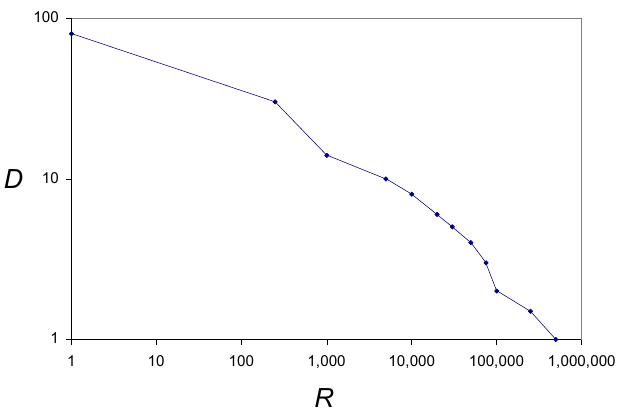}
\end{figure*}

This example contains outlier criteria, $D$, for the variable of interest $V$, classified by another variable,\footnote{The authors would like to thank Gregg R. Diffendal for initial guidance with this example.}  $R$, the reference variable.  $D$ can represent differences or ratios, or more complicated functions, of successive values of $V$.  Normally, the outlier criteria $D$ decline in $R$, so that changes in $V$ take on less importance as $R$ rises.  Table \ref{table2} provides an example of a set of these criteria.
\begin{table}[!ht]
    \centering
\caption{Values of $R$ and $D$}
    \begin{tabular}{|r|r|}
    \hline
        R & D \\ \hline
        500,000+ & 1 \\ \hline
        250,000-499,999 & 1.5 \\ \hline
        100,000-249,999 & 2 \\ \hline
        75,000-99,999 & 3 \\ \hline
        50,000-74,999 & 4 \\ \hline
        30,000-49,999 & 5 \\ \hline
        20,000-29,999 & 6 \\ \hline
        10,000-19,999 & 8 \\ \hline
        5,000-9,999 & 10 \\ \hline
        1,000-4,999 & 14 \\ \hline
        250-999 & 30 \\ \hline
        1-249 & 80 \\ \hline
    \end{tabular}
   \label{table2}
\end{table}
Figure \ref{DRfig} plots $D$ as a function of $R$, where $R$ is taken to be the minimum value of its range and both variables are plotted on logarithmic axes.  Points lying above this curve are considered to be outliers.
Figure  \ref{DRfig} shows that the $(R,D)$ curve is almost a straight line in logarithms.  Therefore, we are led to statistically estimate the equation
\begin{equation}
\log D = a + b\log R.
\end{equation}
The resulting equation produces a straight line intersecting the $(R,D)$ curve, when viewed in logarithms.  Upon rearrangement, we obtain the equation of criticality
\begin{equation}
DR^{-b}=e^a=C.\label{critical}
\end{equation}
An outlier is declared whenever $DR^{-b} > C$. Using the points plotted in Figure \ref{DRfig}, the estimated coefficients are $a = 4.889506$ and $b = -0.33692$.  Substituting these into equation (\ref{critical}) obtains the estimated equation of criticality
\begin{equation}
DR^{0.33692}=132.8879.\label{est}
\end{equation}
This is similar to the loss functions described in Section \ref{uLF1} and estimated in Subsection \ref{preexisting}, with the important difference that the exponent of $R$ is positive.  This occurs because $D$ and $R$ are inversely correlated and $D$ is not an explicit function of $R$.

In our experience, equation (\ref{est}) resulted in missing too many outliers.\footnote{This is an example of the empirical nature of loss functions.}   We decided, instead, to draw a straight line (in logarithms) between the two endpoints, so as to capture all of the previously declared outliers.  The resulting equation (including a slight adjustment to $b$ to ease interpretation) is
\begin{equation}
DR^{\frac{1}{3}}=80.
\end{equation}
Note that $C = 80$, the highest value of $D$.

\part{Data of Mixed Sign\label{part2}}

\section{Introduction to Part \ref{part2}}

We remove the restriction on the direction of the signs in this Part.  This necessitates the creation of much machinery to create useful loss functions.
In Section \ref{LF2}) we specifiy the assumptions and develop the simplest loss function.develops the signed loss function, which incorporates the direction of the difference. In Section \ref{sLF2} we developed the signed loss function and use it to find the optimal parametrization. Unlike Part \ref{part1}, the parametrization depends on the geometry of the loss function, rather than distributions of the data series.

\section{The Loss Function\label{LF2}}

We describe the assumptions and create the simplest loss function, $L(F,B)$, where $F$ is the future value and $B$ is the base period value.  Time can be either chronological or nominal, in which $F$ and $B$ are simply different sets of data. The loss function is the penalty associated with the difference between $F$ and $B$. Roughly speaking, the greater the difference between $F$ and $B$, the greater the loss. After the necessary assumptions are made, the simplest form of $L$ is specified. Restrictions on the values of the parameters of $L$ which make it increase in $B$ for a given average absolute relative difference are then specified.  We give the assumptions and construct the basic form of the unsigned loss function in Subsection \ref{uLF2}. We prove that $(0,0)$ is a continuity point of $L$ under a reasonable assumption in Subsubsection \ref{cp}. Because of the need to deal with the absolute values in $B$ and $F$, we construct the average absolute relative difference and use it to fully specify $L$ in Subsubsection \ref{SigmaL}. We specify the outlier criteria in Subsection \ref{criteria}.   

\subsection{The Unsigned Loss Function\label{uLF2}}

The unsigned loss function $L$ is constructed by making five assumptions. The first assumption is that $L$ is defined everywhere in the real plane $\Re^2$:\newline
{\bf Assumption II.1 (unrestricted domain):} For all $(F, B) \in \Re^2$, $L(F, B)$ is defined and single-valued. \newline
This and Assumption II.2 generalize Assumption I.1.

Assumption II.2 makes $L$ symmetric in the difference $\epsilon$, whether applied to $B$ or $F$:\newline
{\bf Assumption II.2 (symmetry in differences):} $L(B + \epsilon, F)=(B - \epsilon, F)$ and $L(B,F + \epsilon)=(B,F - \epsilon)$
for all $B, F,\epsilon > 0$.\newline
This assumption is relaxed in Section \ref{sLF2}.

A desirable property is that $L$ be symmetric with respect to its arguments. To give a concrete example, we want $L(-1,1000) = L(1000,-1)$. This stated formally as Assumption II.3:\newline
{\bf Assumption II.3 (symmetry in arguments):} $L(B, F) = L(F, B)$.

We now introduce some new notation. Let $X=|F|$ and $Y=|B|$. Let the  loss function $\ell(\epsilon,\Sigma) \equiv L(F,B)$, where $\epsilon = F-B$ and $\Sigma = \Sigma(X,Y)$ is a function such that $\partial\Sigma/\partial X >0$ and
$\partial\Sigma/\partial Y >0$ . Assumption II.3 implies that $\Sigma(X,Y) = \Sigma(Y,X)$, so that $\Sigma$ is symmetric in its arguments. The remaining Assumptions are stated in terms of $\ell$.

Assumption II.4 makes $\ell$ (and $L$) increasing in the difference $\epsilon$:\newline
{\bf Assumption II.4 (monotonicity in difference):} $\partial\ell/\partial\epsilon > 0$ for $\epsilon \ge 0$.\newline
This is identical to Assumption I.2.

Finally, we want $\ell$ to decrease in $\Sigma$.\newline
{\bf Assumption II.5 (monotonicity in arguments):} $\partial\ell/\partial\Sigma < 0$ for all $\Sigma \ge 0$.\newline
This is identical to Assumption I.3.

The simplest function which satisfies Assumptions II.1–II.5 is\footnote{An infinite number of loss functions satisfy Assumptions 1--5. This one is merely the simplest.}
\begin{equation}
	\ell(\epsilon,\Sigma) =
		\left\{
			\begin{array}{cc}
				\epsilon^p \Sigma^q & \Sigma > 0 \\
				0 & \Sigma = 0,
			\end{array}
		\right.\label{LF2P}
\end{equation}
where $p > 0$ and $q < 0$. Note that equation (\ref{LF2P}) is stated in terms of $\epsilon$ and $\Sigma$. The simplest form of $\Sigma$ will be determined in equation (\ref{SigmaEq}) below. Theorem II.1 below shows that setting $\ell(0,0) = 0$ makes $L$ continuous at $(0,0)$, given the choices of $\Sigma,p,q$. This way of defining $\ell(0,0)$ avoids problems associated with division by 0.

\subsubsection{Determination of $\Sigma$ and $L$\label{SigmaL}}

From equation (\ref{LF2P}), it is clear that $\ell(0,\Sigma) > 0$ for all $\Sigma > 0$. We would like to define $\Sigma$ so that whenever either$X >0$ or $Y>0,\Sigma>0.$ We would also like $\Sigma(0,0)=0$. The simplest equation for $\Sigma$ is
\begin{equation}
\Sigma = X + Y = |F| + |B|.\label{SigmaEq}
\end{equation}
Substituting equation (\ref{SigmaEq}) into equation (\ref{LF2P}) and expanding $\Sigma$ obtains the equation for $L$:
\begin{equation}
	L(F,B)=
		\left\{
			\begin{array}{cc}
				|F - B|^p \big(|F|+|B|\big)^q & B \text{ or  }F \neq 0 \\
				0 & B=F=0.
 			\end{array}
		\right.\label{L2ƒ}
\end{equation}
Setting $L(0,0) = 0$ is a convention to simplify its definition.  Subsubsection \ref{cp} shows that this is consistent with $(0,0)$ being a limit point for $L$ when $p + q > 0$.              s

\subsubsection{Proof that $(0,0)$ is a Continuity Point of $L$ when $p + q > 0$\label{cp}}

{\bf Theorem II.1:} $L(0,0) = 0$ is a continuity point of $L(F,B)$ when $p + q > 0$.\footnote{This theorem was proven with the help of ChatGPT.}\newline
{\bf Proof:}  Let $Z = |F| + |B|$ and $r = |F - B|/(|F| + |B|) \in [0,1]$, taking r = 0 at the origin.  Then,
\begin{equation}
L = Z^{p + q} r^p.
\end{equation}
Hence, $0 \le L \le Z^{p + q}$.  Since $p + q > 0$ and $Z \rightarrow 0$ as $F,B \rightarrow 0$, $L \rightarrow 0$ as $F,B \rightarrow 0$ by the squeeze theorem.  Therefore, $(0,0)$ is a continuity point of $L.\Box$

Note that the proof fails for $p + q \le 0$, where no finite limit exists.

\subsubsection{Outlier Criteria\label{criteria}}

See Section \ref{uLF1} for a discussion of outlier criteria and the Lie symmetry that enables the use of $p = 1$. Thus,
\begin{equation}
	L(F,B)=
		\left\{
			\begin{array}{cc}
				|F - B| \big(|F|+|B|\big)^q & B \text{ or  }F \neq 0 \\
				0 & B=F=0.
 			\end{array}
		\right. \label{L}
\end{equation}

A desirable property of the loss function is that it rises in $|F-B|$ for a given average absolute relative difference. The problem lies in defining this difference.  The absolute relative differences with respect to $|F|$ and $|B|$ are $|F-B||F|^{-1}$ and $|F-B||B|^{-1}$, respectively.  Their harmonic mean is
\begin{equation}
	2|F - B|(|F|+|B|)^{-1}.
\end{equation}
Removing the 2 as a nuisance constant, we define the average absolute percentage difference as
\begin{equation}
	|F - B|\big(|F|+|B|\big)^{-1} = \epsilon\Sigma^{-1} \label{aapd}
\end{equation}
for $\Sigma > 0$.  Equation (\ref{aapd}) is the first row of equation (\ref{L}) with $q=-1$.  Choosing $q > -1$ makes the loss function rise in $|F| + |B|$, for a given average absolute relative difference. This is also required by Theorem II.1. We state this as Property II.1:\newline
{\bf Property II.1:} The loss function defined by equations (\ref{L}) increases in $|F| + |B|$ for any given average absolute percentage difference. This is assured whenever $q > -1$.

The reader may note that $q = 0$ turns the first row of equation (\ref{L}) into the absolute values of the differences. Thus, values of $q$ between 0 and $-1$ represent various tradeoffs of absolute differences and average absolute percentage differences. Consider the product of the $r$th power of the absolute difference and the $s$th power of the average absolute relative difference, where $r, s > 0$
\begin{equation}
	|F - B|^r  \left(\frac{|F -B|}{|F|+|B|}\right)^s \label{mult}
\end{equation}
This function is isomorphic to the function $|F - B|\big(|F|+|B|\big)^{r/(r+s)}$. Thus,
these intermediate values of $q$ correspond to an infinite number of pairs $(r, s)$ where $q = r / (r + s)$. Geometrically, the same loss function is generated for all pairs $(r, s)$ lying on the line $s = -(1 - 1/q) r$.

\subsection{The Signed Loss Function\label{sLF2}}

At times, not only the value of the loss function is important, but also the sign of the difference. Different outlier generation processes may manifest themselves by producing predominantly positive or negative differences. We can account for these by creating the signed loss function $S$, which is simply the loss function $L$, multiplied by the signum function of the difference
\begin{equation}
	S(F,B)=
		\left\{
			\begin{array}{cc}
				|F - B| \big(|F|+|B|\big)^q \sgn (F - B) = (F - B) \big(|F|+|B|\big)^q & B \text{ or  }F \neq 0 \\
				0 & B=F=0,
 			\end{array} \label{S2}
		\right.
\end{equation}
where $\sgn x = +1$ for $x > 0$, $0$ for $x = 0$, and $-1$ for $x < 0$.

Using $S$, one can create different critical values for loss, depending on whether the difference is positive or negative. To wit, one can pick $C_+, C_-, C+ \ne -C_-$, such that a pair $(F, B)$ is declared an outlier if either $S(F,B) < C_-$ or $S(F,B) > C_+$. Again, the choice of whether to use $S$ and then use asymmetric critical bounds is an empirical matter.\footnote{The asymmetry need not be limited to the critical values. The signed loss function can incorporate different values of $q$, depending on the sign of the difference.}  The next Subsection investigates $S$ graphically to study the behavior of these loss functions and to suggest choices for $q$.  We can see that the geometry is very important to this choice.

\subsection{Behavior of $S$ and the Choice of $q$\label{plots}}

\begin{figure*}[!hb]
	\caption{$S(F,B),q=0$}\label{II1}
	\includegraphics{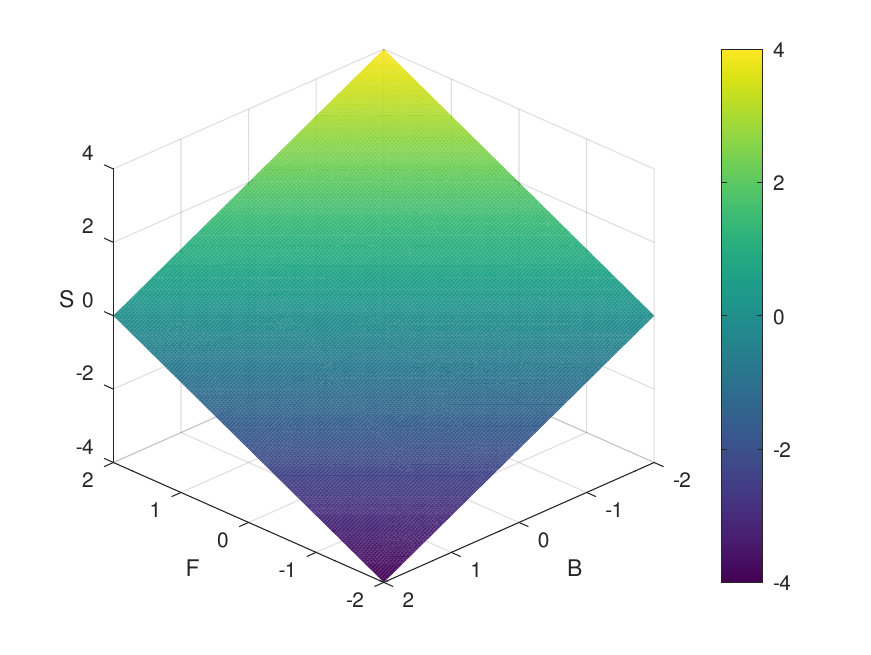}
\end{figure*}

\begin{figure*}
	\caption{$S(F,B),q=-1$}\label{II2}
	\includegraphics{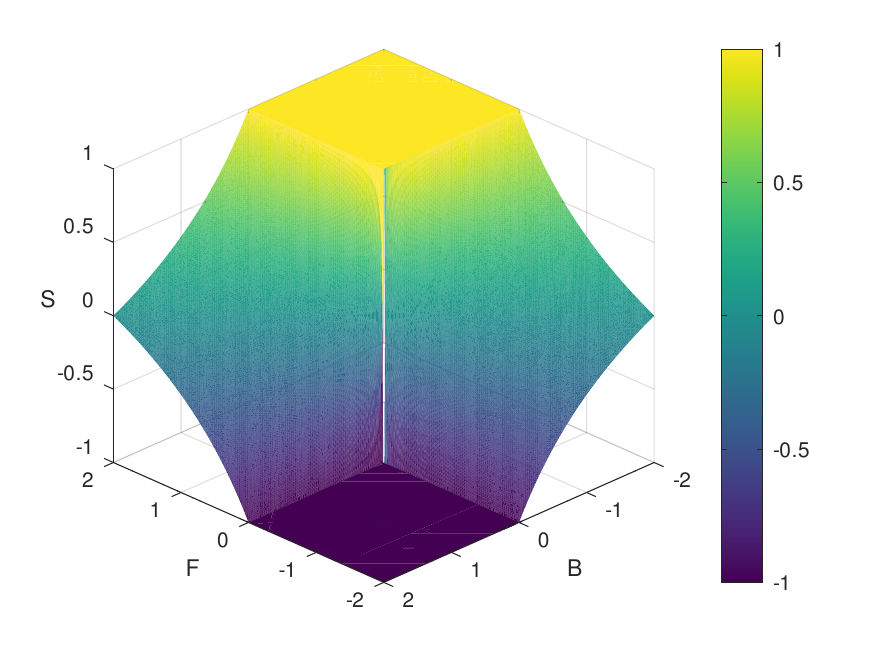}
\end{figure*}

The behavior of $S$ requires that plots of it for various values of $q$ be examined to obtain a reasonable loss function. The limiting functions when $q = 0$ and $q = -1$ are of interest. $q = 0$ implies that $S(F,B) = F – B$. This defines a plane in $\Re^3$, as shown in Figure \ref{II1}.
\begin{center}
	[Insert Figure \ref{II1} about here]
\end{center}
Figure \ref{II2} illustrates $S$ when $q = -1$.
\begin{center}
	[Insert Figure  \ref{II2} about here]
\end{center}
Figure  \ref{II2} shows the lack of a limit at (0,0).  Later Figures have similar artifactual white lines.  Another thing is very apparent from this graph, whenever $B$ and $F$ are of opposite signs, $S(F,B) = \pm\sgn F$. This can be seen by substituting $q = -1$ into equation (\ref{S2}) when $B$ or $F$ is nonzero:
\begin{equation}
	S(F,B)=\frac{F - B} {|F|+|B|} \label{S1}
\end{equation}
Noting that $|x| = x$ when $x > 0$ and $|x| = -x$ when $x < 0$, we can examine the behavior of $S$ when $B$ and $F$ are
of opposite signs. When $F > 0$ and $B < 0$, equation (\ref{S1}) becomes
\begin{equation}
	S(F,B)=\frac{F - B} {|F|+|B|} =\frac{F - B} {F - B} = 1 = \sgn F
\end{equation}
The reader may verify that $S(F,B) = -1 = \sgn F$ when $F < 0$ and $B > 0$. When $F>0$ and $B>0$, the limit becomes path-dependent, lying on $[0,1]$. Finally, when $F<0$ and $B<0$, the limit  lies on $[-1,0]$, with a similar path-dependence. The proofs are omitted. Thus, instead of $S$ having a limit at $(0,0)$, it has a limit set of $[-1,1]$.

\begin{figure*}
	\caption{$S(F,B),q=-.1$}\label{II3}
	\includegraphics{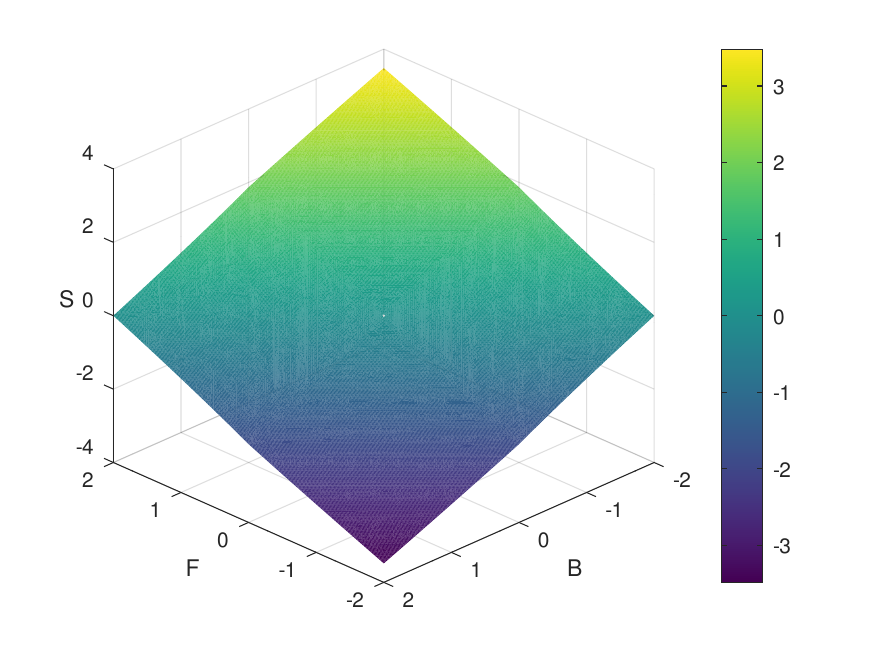}
\end{figure*}

\begin{figure*}
	\caption{$S(F,B),q=-.2$}\label{II4}
	\includegraphics{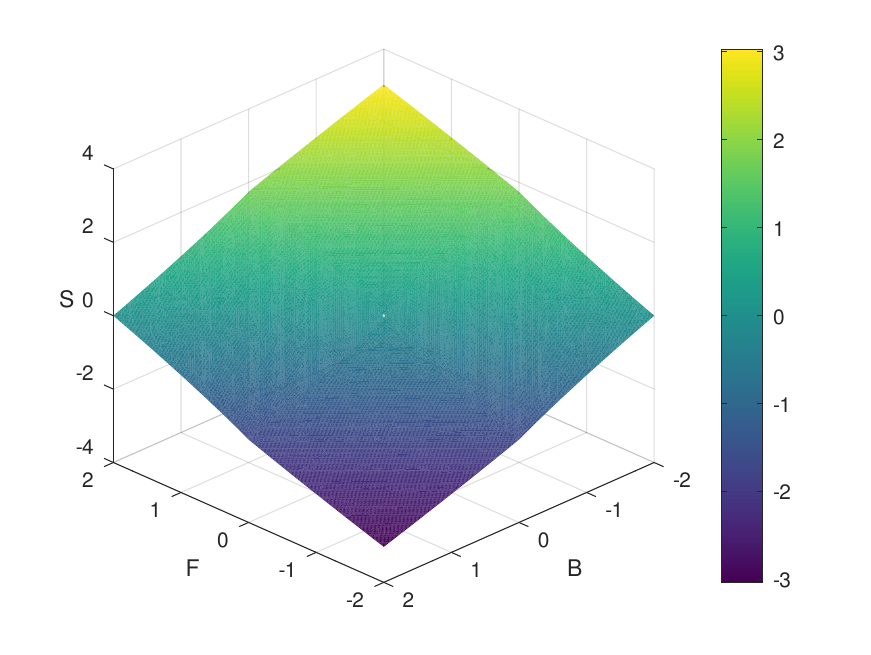}
\end{figure*}

\begin{figure*}
	\caption{$S(F,B),q=-.3$}\label{II5}
	\includegraphics{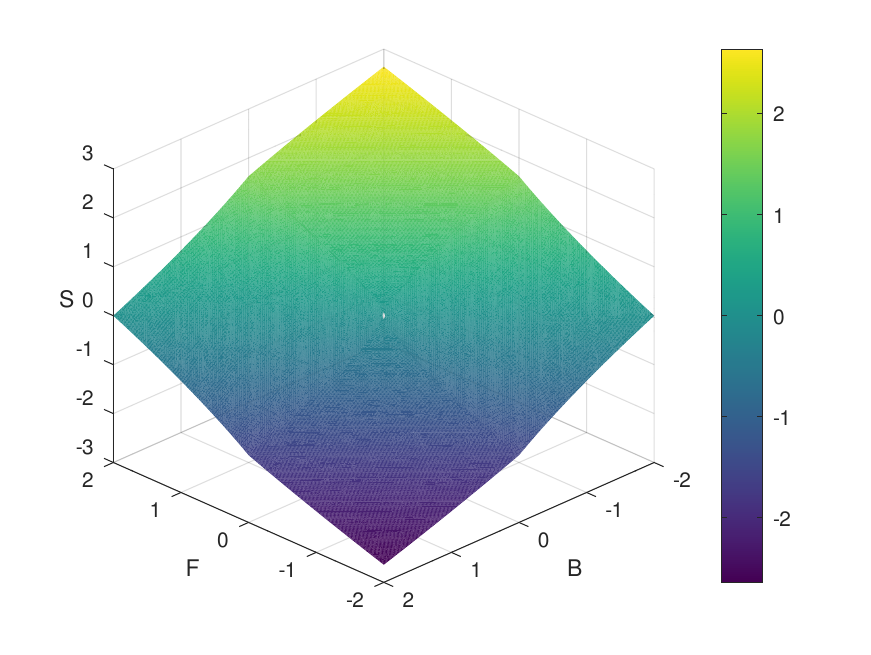}
\end{figure*}

\begin{figure*}
	\caption{$S(F,B),q=-.4$}\label{II6}
	\includegraphics{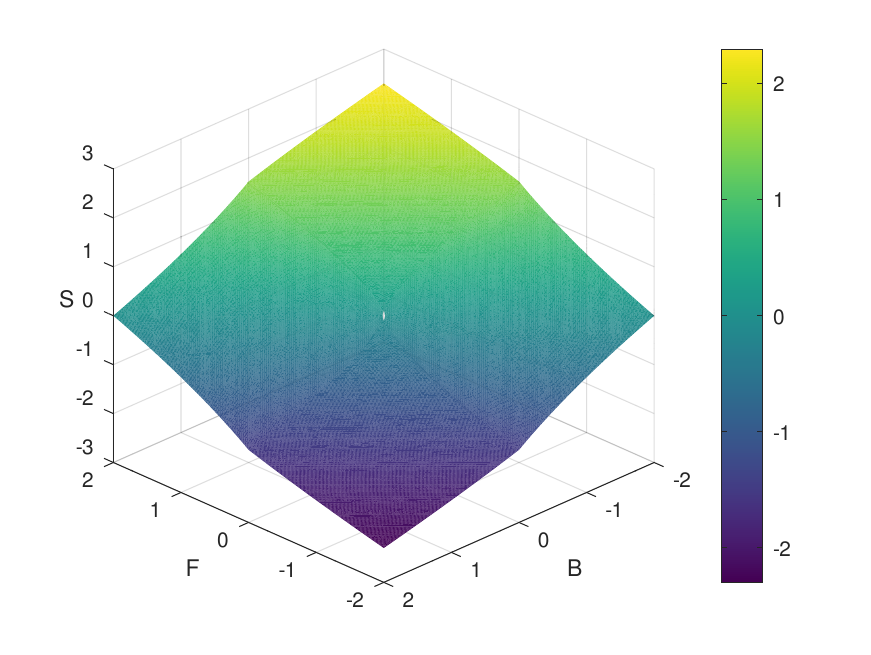}
\end{figure*}

\begin{figure*}
	\caption{$S(F,B),q=-.5$}\label{II7}
	\includegraphics{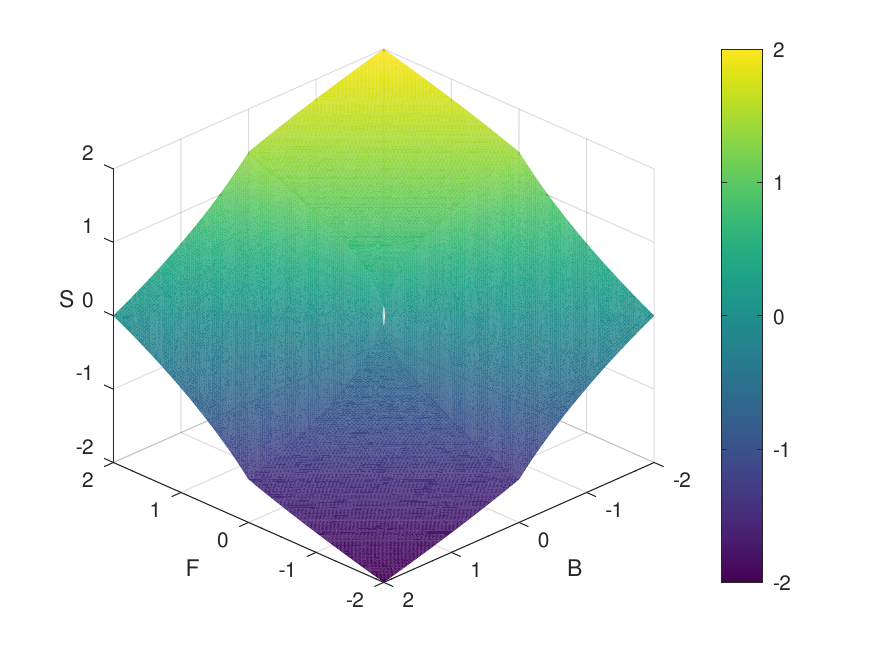}
\end{figure*}

\begin{figure*}
	\caption{$S(F,B),q=-.6$}\label{II8}
	\includegraphics{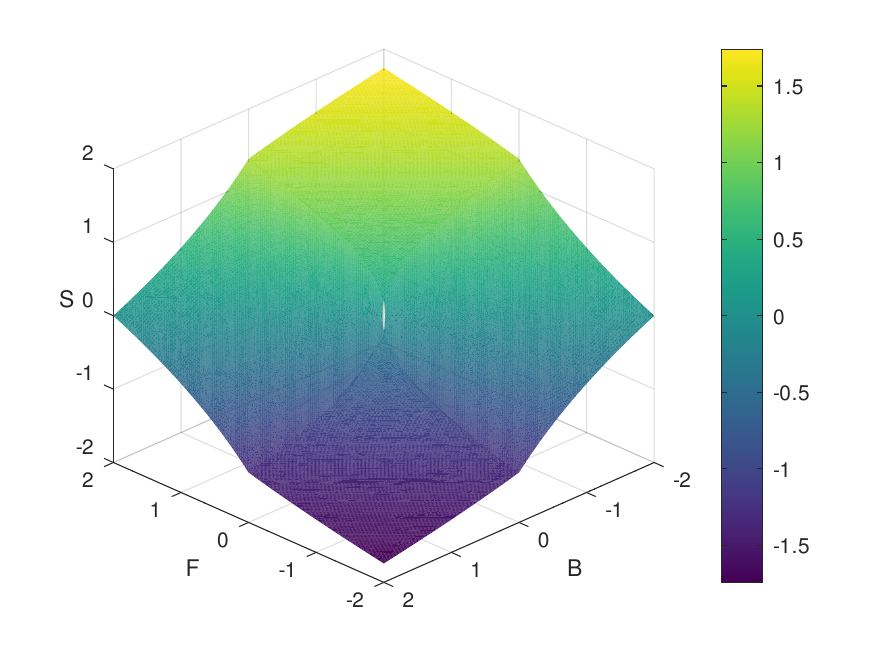}
\end{figure*}

\begin{figure*}
	\caption{$S(F,B),q=-.7$}\label{II9}
	\includegraphics{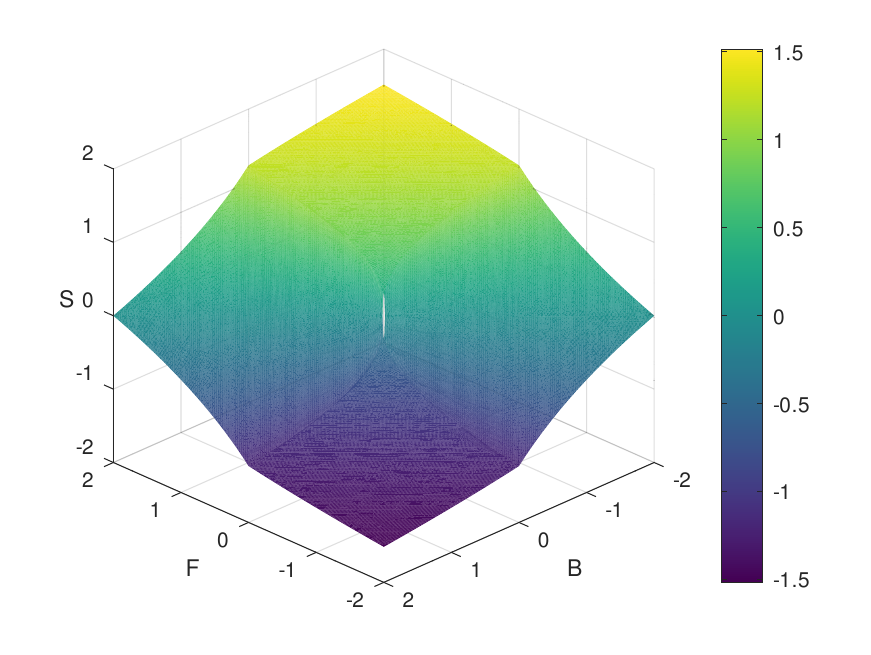}
\end{figure*}

\begin{figure*}
	\caption{$S(F,B),q=-.8$}\label{II10}
	\includegraphics{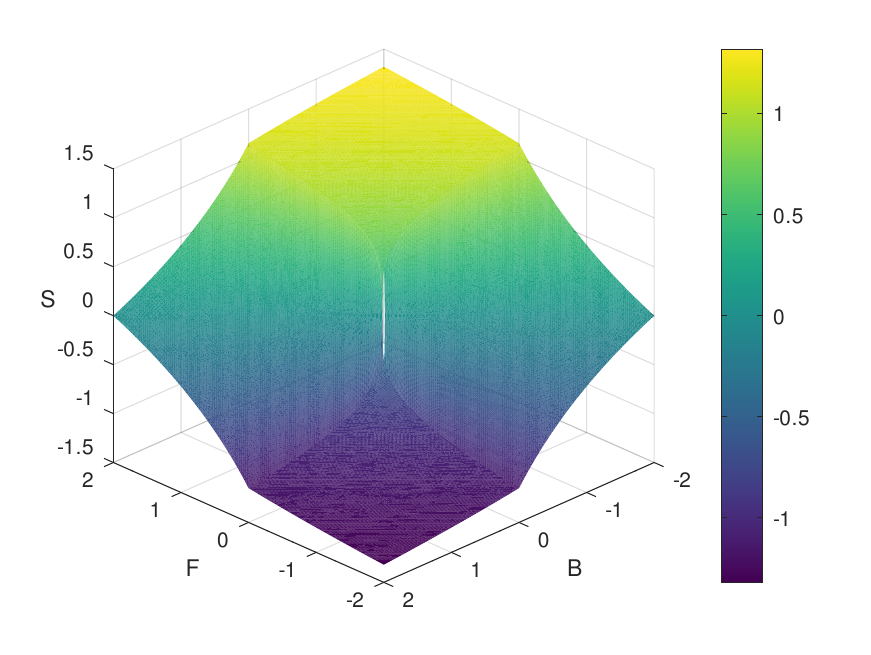}
\end{figure*}

\begin{figure*}
	\caption{$S(F,B),q=-.9$}\label{II11}
	\includegraphics{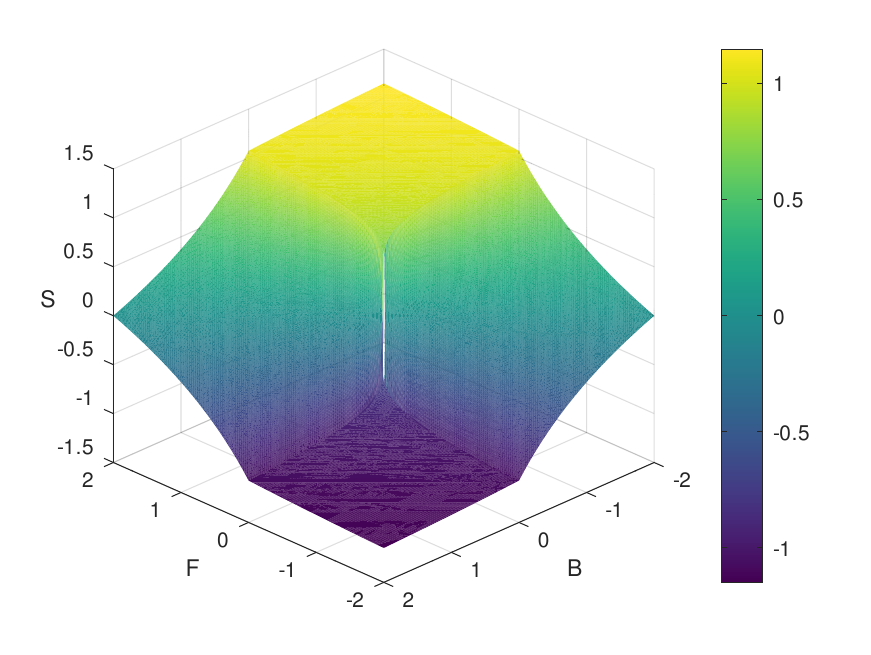}
\end{figure*}

Figure  \ref{II2} shows cusps along the axes. These are also apparent in Figures  \ref{II3}--\ref{II11}, which graph $S$ for $q$ ranging from $-.1$ to $-.9$ in increments of $-.1$, respectively.
\begin{center}
	[Insert Figures \ref{II3}--\ref{II11} about here]
\end{center}
Figures \ref{II3}--\ref{II5} are almost planar, since $q$ is close to 0. However, Figure II.5, $q = -.3$, begins to show the flattening out behavior of Figure II.2. This flattening out increases as $q$ declines to $-.9$ in Figure \ref{II11}. In this Figure, $S \approx \sgn F$ when $B$ and $F$ are of opposite signs.

Given the nonsmooth behavior shown in Figures \ref{II3}--\ref{II11}, the analyst has to decide on a value of $q$ which produces reasonable behavior. It appears that intermediate choices of $q$ are best behaved: these offer a good compromise between simply taking the difference between $F$ and $B$ ($q = 0$) and generating the flatness in $S$ when $q$ approaches $-1$.  Similar to Part \ref{part1}, we recommend starting with $q=-.5$.

\newpage

\section{Figures}

\newpage

\FloatBarrier
\part{Conclusion\label{part3}}

We have extended the concept of outlier detection to panel data.  Outlier detection is of vital importance to many panel data applications, including data quality.  Since no distributional assumptions are made, loss functions are developed to detect these outliers.  Thus, only the possible presence of an outlier can be determined.  Time can either be chronological or nominal.  The direction of the difference can be accounted for. We developed loss functions for nonnegative data in Part \ref{part1}.  We discussed everal applications.  These include showing a general approach to using loss functions, developing loss functions from preexisting outlier criteria, using loss functions with GIS, developing a time-invariant loss function to apply to multiple dates at once and developing a specialized loss function for variables classified by functions of their successive values. In Part \ref{part2}, we dropped the sign restriction to allow any real data.  The geometry restricts the choice of loss function parameter at the cost of allowing the data to inform its choice.

\section*{Authors' Contributions}

Bryan developed several of the ideas.  Coleman performed all derivations and created all text and graphics.

\section*{Supplementary Materials}

The R package \texttt{Outliers} in \url{https://github.com/chuckcoleman/Outliers} contains R code to create the Maps in Part \ref{part1} and Octave code to create the Figures in Part \ref{part2}.

\newpage
\section*{References}

Albert, J. H. (1997).  Bayesian Testing and Estimation of Association in a Two-Way Contingency Table. Journal of the American Statistical Association 92, 685-693.
\newline\newline
Barnett, Vic and Lewis, Toby (1994). Outliers in Statistical Data, 3rd edition, John Wiley \& Sons, New York.
\newline\newline
Bryan, Thomas (1999).  U.S. Census Bureau Population Estimates and Evaluation with Loss Functions. Statistics in Transition 4.
\newline\newline
Coleman, Charles D.  (2025). Loss Functions for Measuring the Accuracy of Nonnegative Cross-Sectional Predictions. \url{https://doi.org/10.48550/arXiv.2505.18130}
\newline\newline
DuMouchel, William (1999). “Bayesian Data Mining in Large Frequency Tables, With an Application to the FDA Spontaneous Reporting System,” The American Statistician 53, 177-188.
\newline\newline
Hoaglin, David C., (1983) “Letter Values: A Set of Selected Order Statistics.”  In, Hoaglin, David C., Frederick Mosteller and John W. Tukey [eds.], Understanding Robust and Exploratory Data Analysis, Wiley, New York.
\newline\newline
Rousseeuw, Peter J., Ida Ruts and John W. Tukey (1999). “The Bagplot: A Bivariate Boxplot,” The American Statistician 53, 382-387.
\newline\newline
Rudas, T., Clogg, C. C. and Lindsay, B. G. (1994). “A New Index of Fit Based on Mixture Methods for the Analysis of Contingency Tables,” Journal of the Royal Statistical Society, Series B, 56, 623-639.
\newline\newline
Tukey, John W. (1977). Exploratory Data Analysis, Addison-Wesley, Reading, Massachussetts.

\end{document}